\documentclass[ aip,prb, reprint]{revtex4-1}
\usepackage{amsmath,amssymb}
\usepackage{graphicx}
\usepackage{float}
\usepackage{tikz}
\usepackage{color}
\usepackage{subfigure}
\usepackage{bm} 

\usepackage[author={Eric Bittner}]{pdfcomment}

\definecolor{mma1}{RGB}{63,60,153}
\definecolor{mma2}{RGB}{135,60,121}
\definecolor{mma4}{RGB}{61,134,99}
\definecolor{mma3}{RGB}{135,123,79} 

\usetikzlibrary{arrows,decorations,backgrounds,snakes}

\begin{document}

\title{Quantum symmetry breaking of exciton/polaritons in metal-nanorod plasmonic array}

\author{Svitlana Zaster}
\address{Department of Chemistry, University of Houston, Houston, TX 77204}
\author{Eric R. Bittner}

\address{Department of Chemistry, University of Houston, Houston, TX 77204}

\author{Andrei Piryatinski}
\address{Theoretical Division, Los Alamos National Laboratory, Los Alamos, NM 87545}

\begin{abstract}
We study the collective, superradiant behavior in the system of emitter-dressed Ag nanorods. Starting from the Drude model for the plasmon oscillations, we arrive at a semi-empirical Hamiltonian describing the coupling between quantized surface plasmon modes and the quantum emitters that can be controlled by manipulating their geometry, spacing, and orientation. Further, identifying the lowest polariton mode as SP-states dressed by excitons in the vicinity of $k=0$, we examine conditions allowing for the polariton quantum phase transition. While the system is formally a 1D array, we show that the polariton states of interest can undergo a quantum phase transition to form a Bose condensate at finite temperatures for physically accessible parameter ranges.  
\end{abstract}
%
\maketitle

\section{Introduction.} 
Spontaneous emission is a powerful probe of the dynamics of a given system.  For a set of isolated emitters, the physics of spontaneous emission is well understood and depends upon the internal details of the specific system.  However, the collective emission from an ensemble of emitters is not so well understood and depends strongly upon the nature of the interactions between the emitters and their surrounding media. Superradiance occurs when the coupling between the emitters and radiation field is sufficiently high that entire system+field must be considered as a collective quantum state and can be described by the classic Dicke model.~\cite{PhysRev.93.99} The Dicke model has been  used to model a wide range of physical systems  from   BEC in optical cavities~\cite{PhysRevLett.105.043001, PhysRevLett.104.130401, Kasprzak:2006mz, PhysRevLett.96.230602} to Hawking radiation from black-holes.~\cite{Nation:2010} A direct realization of the model using a cavity ring-laser was proposed by Dimer {\em et al.}~\cite{PhysRevA.75.013804} Superradiance has also received renewed interest in the field of quantum optics~\cite{PhysRevA.75.013804} and nanotechnologies.~\cite{Agranovich:1997nx,springerlink:10.1134/1.1130691,PhysRevA.82.023827}
It also underlies the physics behind the formation of exciton/polariton Bose condensates in thin-film micro cavities.
 \cite{Byrnes:2010uq,Keeling:2007fk,Amo:2009fk, Kasprzak:2006mz, bittner:034510,Kasprzak:2008mi,Zoubi:2006rb,assmann10022011, michetti:195301, DavidSnoke11152002, Richard:2005tw,Wertz:2010aa, wertz:051108,PhysRevB.72.125335,PhysRevB.81.081307, D.-D.-C.-Bradley:1998xw}

Quantum plasmonics is a rapidly emerging field of nanotechnologies establishing control of light quantum properties via its interaction with nanoscale metallic structures showing surface-plasmon (SP) response.~\cite{TameMS:2013} Typically the SP modes are excited via near-field interactions with quantum emitters such as atomic impurities or semiconductor nanostructures. The coupling between the quantum emitters and SP modes plays a key role in controlling the  quantum properties of the photons. Previous theoretical investigations indicate that strong coupling between an ensemble of quantum emitters and a SP mode can result in superradient emission.~\cite{Pustovit:09,Pustovit:10} Such calculations utilize a semiclassical approach and provide the correct trends in the emission line-shape with the number of emitters allowing for identification of the super- and sub-radiant exciton-SP modes. However, investigation of cooperative thermodynamic properties leading to spontaneous symmetry breaking associated with the superradiant regime requires a fully quantum mechanical approach.     

\begin{figure}[b]
\begin{center}
\begin{tikzpicture}[scale = 1.4]
\draw[->,color=black](0,0)--(4,0); \draw (4.1,0)  node {$x$};
\draw[->,color=black](0,0)--(0,1.5);  \draw (0,1.6)  node {$y$};
\draw[->,color=black](0,0)--(-1,-.75);  \draw (-1.1,-0.8)  node {$z$};
\draw[color=black!20](1,0)--(0,-.75);  
\draw[color=black!20](2,0)--(1,-.75);  
\draw[color=black!20](3,0)--(2,-.75);  
\draw[color=black!20](4,0)--(3,-.75);  
\draw[color=black!20](0,0)--(1,0.75);  
\draw[color=black!20](1,0)--(2,0.75);  
\draw[color=black!20](2,0)--(3,.75);  
\draw[color=black!20](3,0)--(4,.75);  
\draw[color=black!20](4,0)--(5,.75);  
\draw[color=black!20](-1,-0.75)--(3,-0.75);
\draw[color=black!20](-1,0.75)--(5,0.75);
\draw[color=black!20](0,0.75)--(-1,0);
\draw [snake=snake,segment amplitude=2mm,draw=mma2] (-1,0) -- (4.,0);
\draw (3,0.6)--(3,0);
\filldraw[fill=blue!20,draw=blue!50!black,rotate around={-60:(0,0)}] (0,0) ellipse (.2cm and 1.0cm);
\draw[dashed] (0,0) ellipse (0.15cm and 0.2cm);
\filldraw[fill=red!20,draw=red!50!black] (0,.6) circle (.1cm);
\filldraw[fill=blue!20,draw=blue!50!black,rotate around={-60:(3,0)}] (3,0) ellipse (.2cm and 1.0cm);
\filldraw[fill=red!20,draw=red!50!black] (3,.6) circle (.1cm);
\draw[dashed] (3,0) ellipse (0.15cm and 0.2cm);
\draw (1.5,0.8) node {$J_{sp}$};
\draw[<->] (.2,0.4) arc (100:80:220pt);
\draw (-0.6,0.4) node {$\lambda_{nk}$};
\draw[<->] (-0.2,0.6) arc (150:210:20 pt);
\draw (1.5,-.4) node {$\hat\psi_{k}$};
\draw[->] (-.54,-0.25)--(.46,0.25); 
\draw[->] (2.54,-0.25)--(3.46,0.25); 
\draw (3.7,0.3) node {$\bm p_{sp}$};
\draw (3.5,.8) node {$\bm \mu_{qe}$};
\draw[->] (2.78,0.5  ) -- (3.22,0.7);
\draw[->] (-0.22,0.5  ) -- (.22,0.7);
%
\end{tikzpicture}
\end{center}
\caption{Repeated unit of a one-dimensional periodic structure of metal nanorods (blue spheroieds) lying in the $xz$ plane with period $a$ along x-axis. Angle $\theta$ between spheroid's major axis and x-axis defines orientation of the SP-mode dipoles $\bm p_{sp}$. Each nanorod is dressed by a quantum emitter (red disk) located a distance $d$ above the nanorod and  characterized by a transition dipole moment $\bm\mu_{qe}$ held parallel to ${\bm p}_{sp}$. The wavy curve represents a collective plasmon mode $\hat\psi_{k}$ due to near-field coupling between the nanorods, $J_{sp}$ (Eq.~\ref{Jsp-sprd}). The emitters  are coupled to the collective SP mode via $\lambda_{nk}$ (Eq.~\ref{eq:lambda}). 
}
\label{Fig-gm}
\end{figure}
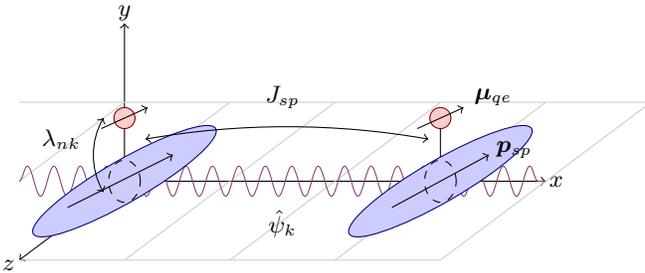

To address this problem, we examine the on-set of collective excitonic states in one-dimensional periodic arrays of  Ag nanorods dressed by dipole emitters as sketched in Fig.~\ref{Fig-gm}. The near-field coupling between localized SP modes within the nanorods and the dipole emitters gives rise to a collective exciton-SP mode. Using a quantum mechanical model that accounts for the explicit interactions between the quantum emitters and SP, we construct a Dicke Hamiltonian.~\cite{PhysRev.93.99,PhysRevLett.104.130401} In contrast to conventional Dike model where the quantum emitters are coupled to radiative cavity mode, our model couples the quantum emitters via near-field quasi-static Coulomb interactions. Within this model, the coupling between the emitters and the SP modes enhance the Coulomb couplings between emitters that can be controlled by manipulating the aspect-ratio of the nanorods and their orientation. Since the transverse radiative potential is not included in the near-field regime, our model is free of an issues raised in the literature regarding the gauge invariant form of the Hamiltonian that preserves Thomas-Reiche-Kuhn sum rule.~\cite{Rzazewski:1975} By analyzing the thermodynamics of this collective system, we predict that symmetry breaking leading to the collective, superradiant regimes.  

The paper organized as follows. In Secs.~\ref{Sec:SP-model} we derive second-quantized SP model accounting for the local field interactions. In
Sec.~\ref{Sec:SPX-model}, the Hamiltonian for the near-field coupled collective states of SP and quantum emitters and their mutual interactions is introduced. Based on this Hamiltonmian, in Sec.~\ref{Sec:QSB} we evaluate SP and exciton polariton modes and identify the lower-polariton mode as exciton-dressed SP states in the vicinity of $k=0$. Further examination of the conditions for the symmetry breaking for this mode demonstrates the possibility of quantum-phase transition into the superradiant state at finite temperature. Sec.~\ref{Sec:Conc} concludes the paper. 

\section{Interacting SP model}
\label{Sec:SP-model}

We shall begin by assuming that the metal nanoparticles interact  via quasi-electrostatic interactions, giving rise to a collective surface-plasmon (SP) mode. Each nanoparticle is dressed by a single dipole-emitter, which in turn are coupled both to each other via near-field interactions allowing an optical excitation to be exchanged between emitters and to the plasmon mode.  For the sake of simplicity, we further assume that the metal nanoparticles, i.e., the nanorods, are strongly anisotropic and can be characterized by a single SP mode with the energy $\omega_{sp}$. This mode is polarized in the $(x,z)$-plane along the unit vector ${\bf e}_{sp}$ as illustrated in Fig.~\ref{Fig-gm}. 
  We also assume that the quantum emitter transition dipole is polarized in the same plane along the unit vector ${\bf e}_{qe}$ and parallel to the 
 major-axis of the nanorod.   This allows us to define the SP  and the quantum emitter transition dipole operators as
\begin{eqnarray}
\label{musp}
\hat{\bm p}_n &=& {\bf e}_{sp}p_{sp}\left(\hat \psi_n^\dag+\hat \psi_n\right),
\\\label{mux}
\hat{\bm\mu}_n &=& {\bf e}_{qe}\mu_{qe}\left(\hat\sigma^+_n+\hat\sigma^-_n\right),
\end{eqnarray}
respectively. Here $\{\hat\psi_{n} ,\hat\psi_{n}^{\dagger}\}$ are boson operators for the SPs. The quantum emitters are approximated as two-level systems and described by SU(2) spin operators $\hat\sigma_n^{i}$ with $i=\pm, z$. The  quantity, $\mu_{qe}$, is the transition dipole matrix element between the ground and the excited states of each quantum emitter. Unless different values are specified, below we set quantum emitter transition dipole strengths to $\mu_{qe}= 75$~Debye. Such a large value can be expected in epitaxially grown semiconductor quantum dots.~\cite{StievaterPRL-2001}.  The amplitude of the metal nanoparticle SP dipole is
\begin{eqnarray}
\label{p-sp}
p_{sp}=\left(\frac{\omega_{sp}\alpha_{sp}}{2}\right)^{1/2}.
\end{eqnarray}
This depends on the SP energy, $\omega_{sp}$  and the nanoparticle polarizability, $\alpha_{sp}$ which we now derive.

To describe anisotropic SP response of the metal nanoparticles forming the array, we define them to be prolate spheroids and set the spheroid major axis to be oriented in the direction of the unit vector ${\bf e}_{sp}$ constrained to the  $(x,z)$-plane. We further denote the spheroid  major (minor) radius by $a_0$ ($b_0$), the distance to the spheroid focal point as $f_0=\sqrt{a_0^2-b_0^2}$ and the inverse eccentricity as $\eta_0=[1-(b_0/a_0)^2]^{-1/2}$. 

For large enough values of the aspect ratio $a_0/b_0$, one can consider the spheroidal nanoparticles as the nanorods whose SP mode oscillating along the major axis couples strongly to the local electric fields whereas the contribution of the SP oscillations along the minor axis can be neglected. In this approximation, dipole moment of the spheroidal nanorod located at the $n$-th site induced by the local electric field $E_{n}$ can be represented as 
\begin{eqnarray}
\label{p-sph}
 p_{n}(\omega) = \alpha(\omega)E_{n}
\end{eqnarray}
where the nanorod polarizability 
\begin{eqnarray}
\label{pol-spher}
 \alpha(\omega)=\frac{4\pi\varepsilon_0 f_0^3}{3}\frac{\epsilon(\omega)-1}
 				{[Q_1(\eta_0)/\eta_0]\epsilon(\omega)-{Q_1'}(\eta_0)},
\end{eqnarray}
depends on the spheroid dielectric function $\epsilon(\omega)$ and the geometric factors given in terms of the Legendre polynomial  of the second kind, $Q_1(z)=(z/2)\ln[(z+1)/(z-1)]-1$, and its derivative, $Q_1'(z)$. 
  
To quantize the plasmon modes, the nanorods are described by the Drude dielectric function  
\begin{eqnarray}
\label{e-Drude}
 \epsilon(\omega)=\epsilon_\infty-\frac{\omega_p^2}{\omega(\omega+i\gamma)},
\end{eqnarray}
where $\epsilon_\infty$ is the dielectric constant due to the high frequency response of the metal ion core and the second term describes the response of the metal electron gas characterized by the bulk plasma frequency, $\omega_p$, and damping rate, $\gamma$. For the numerical calculations  we use parameters entering Drude dielectric function obtained by Johnson and Christy of the SP response for bulk Ag.~\cite{johnson72}  Specifically, we set the high frequency dielectric constant of the ion core to $\epsilon_\infty=3.7$, the electron plasma  frequency to $\omega_p= 9.1$~eV, and the damping constant to $\gamma = 18$~meV. The environment dielectric constant is set to unity. 

Substituting Eq.~(\ref{e-Drude})  into Eq.~(\ref{pol-spher}), we recast Eq.~(\ref{p-sph}) to the following form
\begin{eqnarray}
\label{nppzw}
p_{n}(\omega)=\left[\alpha_\infty+\alpha_{sp}\frac{\omega^2_{sp}}{\omega(\omega+i\gamma)-\omega^2_{sp}}\right]E_n(\omega),
\end{eqnarray}
where 
\begin{eqnarray}
\label{pol-core}
 \alpha_\infty &=&\frac{4\pi\varepsilon_0 f_0^3}{3}\frac{\epsilon_\infty-1}
 				{[Q_1(\eta_0)/\eta_0]\epsilon_\infty-{Q_1}'(\eta_0)},
\\\label{pol-sp}
\alpha_{sp} &=&\frac{4\pi\varepsilon_0 f_0^3}{3}\frac{[Q_1(\eta_0)/\eta_0]-{Q_1}'(\eta_0)}
 				{[Q_1(\eta_0)/\eta_0]\epsilon_\infty-{Q_1}'(\eta_0)},		
\end{eqnarray}
are the higher frequency ion core and the low frequency electron gas polarizabilities, respectively. According to Eq.~(\ref{pol-sp}), the latter quantity prefactors the resonant response of the electron gas on the SP frequency of the nanorod,  
\begin{eqnarray}
\label{w-sp}
 \omega_{sp}^2 = \frac{[Q_1(\eta_0)/\eta_0]~\omega^2_p}{[Q_1(\eta_0)/\eta_0]\epsilon_\infty-{Q_1}'(\eta_0)}.		
\end{eqnarray}
\noindent
\begin{widetext}
In the time domain Eq.~(\ref{nppzw}) reads
\begin{eqnarray}
\label{nppzt}
&~&p_{n}(t)=\alpha_{sp}\omega_{sp}^2
	\int^t_{-\infty}dt'\frac{\sin\left[\sqrt{\omega^2_{sp}-\gamma^2/4}~(t-t') \right]}
	{\sqrt{\omega^2_{sp}-\gamma^2/4}}e^{-\gamma (t-t')/2}E_n(t'),
\end{eqnarray}
\end{widetext}
which is solution for the equations of motion of a driven harmonic oscillator  
\begin{eqnarray}
\label{psp-eqm}
\ddot p_{n}(t)+\gamma\dot p_{n}(t)+\omega_{sp}^2 p_{n}=\alpha_{sp}\omega_{sp}^2E_n(t).
\end{eqnarray}

Now we introduce an effective SP coordinate, $u_n$, and conjugate momentum $\pi_n$. By neglecting the damping term in Eq.~(\ref{psp-eqm}), we further introduce effective SP Hamiltonian 
\begin{eqnarray}
\label{Hsp-clas}
 	H_{SP} = \sum_{n=1}^N\left(\frac{\pi_n^2}{2 m_{sp}}+m_{sp}\omega_{sp}^2\frac{u_n^2}{2}\right) - \sum_{n=1}^N p_{n}E_n,
\end{eqnarray}
where $p_{sp}=eu$ with $e$ being electron charge. By writing equation of motion for $u_n$ and comparing it with Eq.~(\ref{psp-eqm}), one can easely identify the SP mass as $m_{sp}=e^2/\alpha_{sp}\omega_{sp}^2$. Second quantization of the SP momentum and coordinate
\begin{eqnarray}
\label{pin-psi}
\hat \pi_n&=&i\sqrt{\frac{\omega_{sp} m_{sp}}{2}}\left(\hat\psi_n^\dag-\hat\psi_n\right),
\\\label{un-psi}
\hat u_n&=&\sqrt{\frac{1}{2\omega_{sp}m_{sp}}}\left(\hat\psi_n^\dag+\hat\psi_n\right),
\end{eqnarray}
results in the following Hamiltonian (here and below $\hbar = 1$)
\begin{eqnarray}
\label{Hsp-quant}
 	\hat H_{SP} = \omega_{sp}\sum_{n=1}^N\hat\psi_n^\dag\hat\psi_n - \sum_{n=1}^N \hat {\bm p}_{n}\cdot\hat{\bm E}_n,
\end{eqnarray}
with the dipole operator defined by Eq.~(\ref{musp}) and the local electric field $\hat{\bm E}_n(t)$ to be a superposition of the electric fields produced by the metal nanoparticles at sites $m\neq n$ and quantum emitters at all sites. Specifically,
\begin{eqnarray}
\label{En-local}
 	\hat{\bm E}_n &=& -\frac{3}{8\pi\varepsilon_0 f_{0}^3}
	\sum_{m=1}^N\left[1-\delta_{nm}\right] {\bm T}([m-n]a,0)\hat p_m
\\\nonumber 
 	 &-&\frac{3 }{8\pi\varepsilon_0 f_{0}^3}
	\sum_{m=1}^N \left(\hat{\bm \mu}_m\cdot{\bm T}([m-n]a,d)\right){\bf e}_{sp}.\;\;\;\;\;\;\;\;
\end{eqnarray}
Here, the dimensionless vector determining direction of the local electric field induced by the spheroid polarization along its major axis is
\begin{eqnarray}
\label{T-local}
 {\bm T}(x,y,z)=\bm\nabla\left[ z' \ln\left(\frac{r_1'+r_2'+2f_0}{r_1'+r_2'-2f_0}\right)
	-r_1'+r_2' \right],\;\;\;\;
\end{eqnarray}
where
\begin{eqnarray}
\label{r12-prime}
 r'_{1,2}=\sqrt{{x'}^2+y^2+(z'\pm f_0)^2},
\end{eqnarray}
and
\begin{eqnarray}
\label{XZ'toXZ}
\left(\begin{array}{c}
 x'\\z'
\end{array}\right)=
\left(\begin{array}{cr}
 ({\bf e}_{sp}\cdot{\bm z})&-({\bf e}_{sp}\cdot{\bm x}) \\
 ({\bf e}_{sp}\cdot{\bm x})& ({\bf e}_{sp}\cdot{\bm z})
\end{array}\right)
\left(\begin{array}{c}
 x\\z
\end{array}\right).
\end{eqnarray}

After assuming that $f_0\ll r_{1,2}$, Eq.~(\ref{En-local}) can be written in the following dipole-filed form
\begin{eqnarray}
\label{En-local-dipole}
 	\hat{\bm E}_n &=& -\sum_{m=1}^N\left[1-\delta_{nm}\right]
	\frac{\hat{\bm p}_m-3\bar{\bm r}_{nm}(\bar{\bm r}_{nm}\cdot\hat{\bm p}_m)}{4\pi\varepsilon_0 r_{nm}^3}
\\\nonumber&-&	
	\sum_{m=1}^N
	\frac{\hat{\bm \mu}_m-3\bar{\bm R}_{nm}(\bar{\bm R}_{nm}\cdot\hat{\bm \mu}_m)}{4\pi\varepsilon_0 R_{nm}^3}.
\end{eqnarray}
Here $r_{nm}$ ($R_{nm}$) is the distance between metal nanoparticle on site $n$ and a metal nanoparticle (quantum emitter) on site $m$, and $\bar {\bm r}_{nm}=\bm r_{nm}/r_{rm}$ ($\bar {\bm R}_{nm}=\bm R_{nm}/R_{rm}$).

\section{Coupled SP-exciton Hamiltonian}
\label{Sec:SPX-model}
\begin{widetext}
Having defined our terms, we can now move on to define the total Hamiltonian describing the physical system. For this we extend Eq.~(\ref{Hsp-quant}) as 
\begin{eqnarray}
\label{Hsp-quant2}
\hat H_{sp-x} &=&
\sum_{k}\Omega_{sp}(k)\hat\psi_{k}^{\dagger}\hat\psi_{k} + 
\frac{1}{\sqrt{2N}}\sum_{n}\sum_k\left(\hat\sigma_{n}^{+}+\hat\sigma_{n}^{-}\right)\left(\lambda_{nk}\hat\psi_{k}^{\dagger}
+\lambda_{n k}^*\hat\psi_{k}\right)+\hat H_{XY}.  
\end{eqnarray}
where the first term describes interaction free SP mode with the energies $\Omega_{sp}(k)$. Field operators $\hat \psi_k$ and  $\hat \psi^\dag_k$ obey Bose commutation relation and describe collective SP modes of the nanorod array characterized by the wave vector $k$. The second term describes coupling of the SP mode to quantum emitters with the ground-excited state transitions described by the spin operator $\hat\sigma_n^{\pm}$ with $n=1,2,\dots N$ being the array site index.  The last term, $\hat H_{XY}$, describes the quantum emitters and its explicit form will be specified below. 
\end{widetext}
\subsection{Coupling between neighboring metal nanorods}

The first term in the Hamiltonian given by Eq.~(\ref{Hsp-quant2}) is derived by starting with the effective SP Hamiltonian (Eq.~(\ref{Hsp-quant})) in which the second term is calculated using the local field coupling between the nearest neighbor nanorods (i.e., the first term in Eq.~(\ref{En-local}) with $m=n\pm 1$) and second quantized form of the SP dipole (Eq.~(\ref{musp})). This results in the following near-field interaction energy between the nearest neighbor nano rods 
\begin{eqnarray}
\label{Jsp-sprd}
J_{sp}= \frac{3p_{sp}^2}{8\pi\varepsilon_0f_0^3}({\bf e}_{sp}\cdot\bm T(a,0)),
\end{eqnarray}
where $f_0=\sqrt{a_0^2-b_0^2}$ is the distance to the focal point of the spheroid measured from its center and vector $\bm T(x,z)$ given by Eqs.~(\ref{T-local})--(\ref{XZ'toXZ}). $a$ is the distance between centers of nearest neighbor nanorods. In the limit $f_0\ll r_{1,2}$ describing the nanorod separation significantly exceeding their size, Eq.~(\ref{Jsp-sprd}) recovers the spherical dipole-dipole interaction term
\begin{eqnarray}
\label{Jsp-dd}
J_{sp}^{dd}= \frac{p_{sp}^2}{4\pi\varepsilon_0 a^3}\left(1-3\left({\bf e}_{sp}\cdot{\bm x}\right)^2\right).
\end{eqnarray}
Note that this expression directly follows from the first term in the asymptotic expression of the local field given by Eq.~(\ref{En-local-dipole}). 
\begin{figure*}
\subfigure[]{\includegraphics[width=0.62\columnwidth]{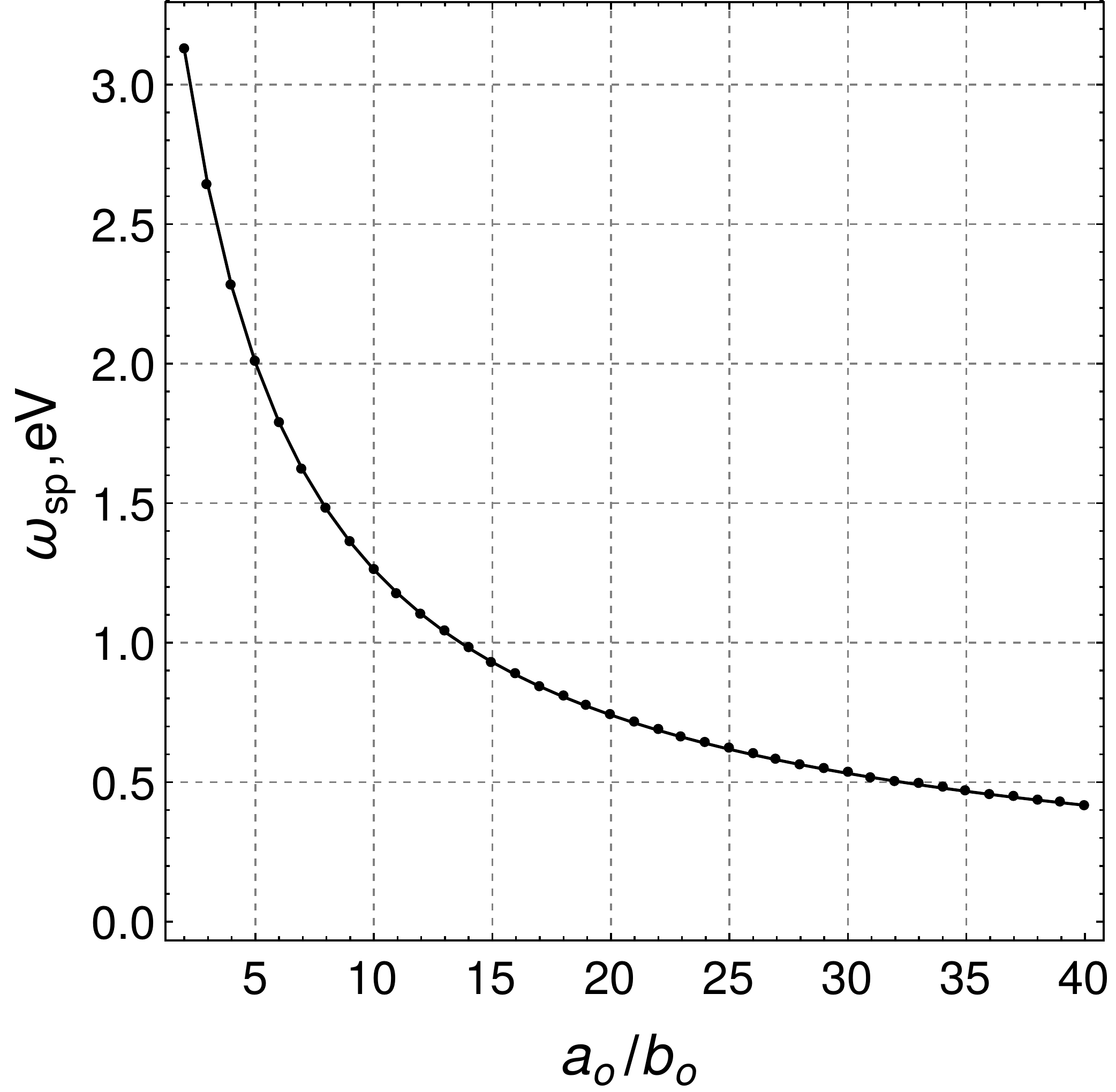}}
\subfigure[]{\includegraphics[width=0.62\columnwidth]{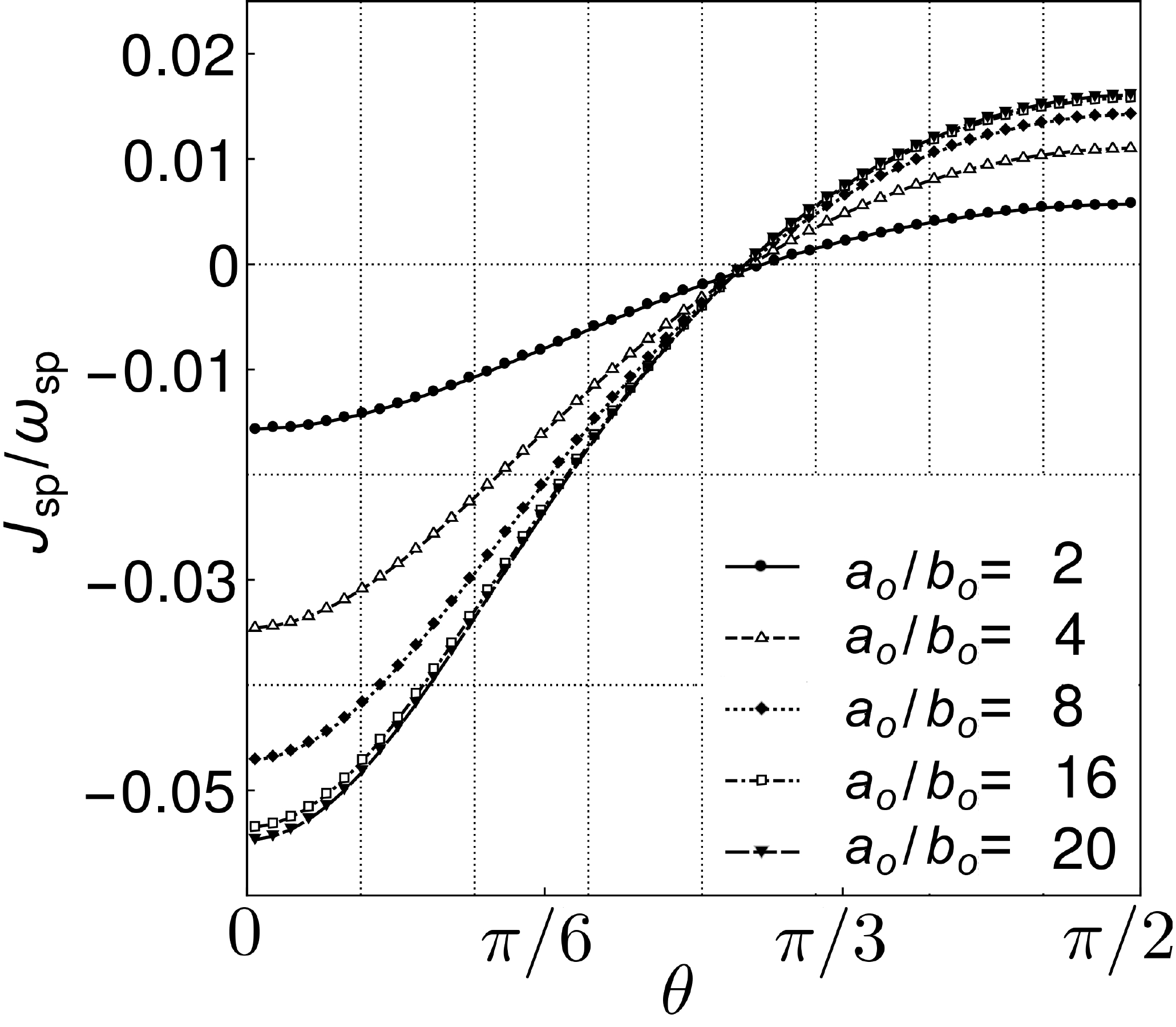}}
\subfigure[]{\includegraphics[width=0.62\columnwidth]{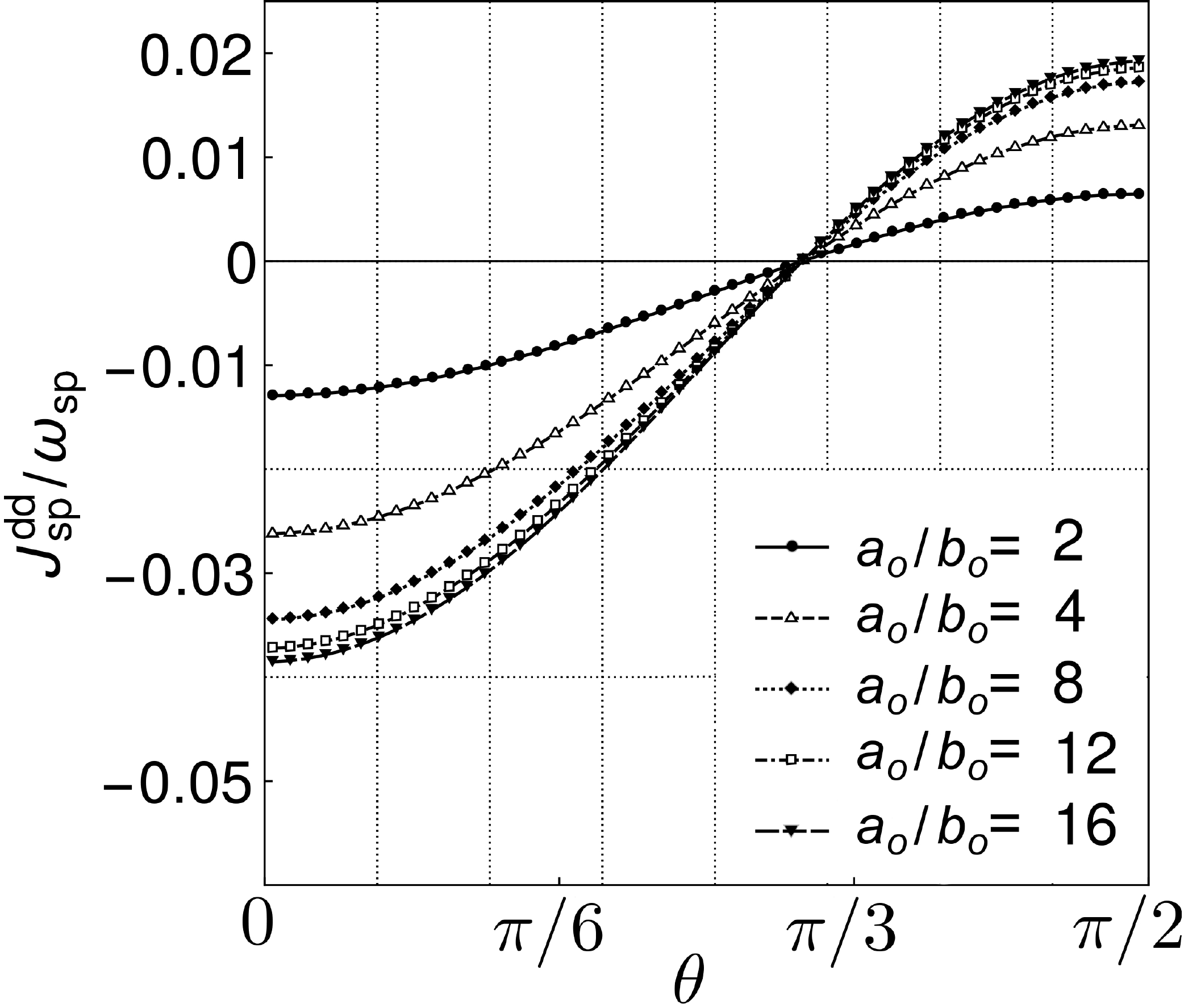}}
\caption{(a) The SP energy, $\omega_{sp}$, as a function of nanorod aspect ratio $a_{o}/b_{o}$. 
(b) Normalized near-field coupling between  nanorods (Eq.~(\ref{Jsp-sprd})) as a function of tilt-angle $\theta$ 
calculated for various aspect ratios $a_{o}/b_{o}$ and $b_{o}=2$~nm . (c) For comparison, normalized  
dipole-dipole coupling energy calculated for the same parameters according to Eq.~(\ref{Jsp-dd}) is provided.  
}\label{Fig-Jvalues}
\end{figure*}
Furthermore, neglecting the coupling terms between the next to the nearest neighbor nanoparticles and higher, we diagonalized the SP terms by performing its Fourier transformation.  This results in the first term in Eq.~(\ref{Hsp-quant2}) with the SP energy dispersion  
\begin{eqnarray}
\label{wspk}
\Omega_{sp}(k)=\omega_{sp}+2J_{sp}\cos(k), \label{sp-dispersion}
\end{eqnarray}
where $k=-\pi,\dots,-\pi$.  

According to Eq.~(\ref{wspk}) the plasmonic band is determined by the values of $\omega_{sp}$ and $J_{sp}$. Fig.~\ref{Fig-Jvalues}a shows the variation of the surface plasmon energy $\omega_{sp}$ with increasing nanorod aspect ratio $a_0/b_0$, where $a_0$ ($b_0$) is spheroid major (minor)  radius.  By simply varying the major radius, one can tune this frequency over a wide range spanning the near infrared through ultraviolet spectrum. In Fig.~\ref{Fig-Jvalues}b, we show the variation in the near-field coupling between emitters (Eq.~(\ref{Jsp-sprd})) for various nanorod aspect ratios $a_0/b_0$. The position of the SP band edge within the Brillouin zone depends on the sign of $J_{sp}$. According to Fig.~\ref{Fig-Jvalues}b, the sign of $J_{sp}$ can be varied from negative to positive by changing the polarization of the SP mode from the ${\bm x}$ to ${\bm z}$ directions. The cross-over between  negative and positive values of $J_{sp}$ occurs at an angle close to the $\theta = 54.7^{\circ}$. For the cases shown here, the distance of separation between the centers of the nanorods is held fixed at $a = 2 a_{o} + \delta a$ and the end-to-end distance $\delta a=1$~nm ensures that even at $\theta = 0$ (linear aligned rods), the rods do not touch or overlap.  For more closely packed rods (not shown in the plot), near-field effects become increasingly important and the cross-over from negative to positive values of $J_{sp}$ occurs at smaller angles. Comparison of the exact (Eq.~(\ref{Jsp-sprd})) and dipole-dipole (Eq.~(\ref{Jsp-dd})) terms is provided in panels b and c of  Fig.~\ref{Fig-Jvalues}. The plot shows that the deviation from the dipole-dipole approximation is weak and becomes observable for small angles and large aspect ratios.

\subsection{Exciton-SP interaction term}

By performing the substitution of the expression for the local electric field produced by quantum emitters (i.e., the second term in Eq.~(\ref{En-local}) into the second term of the effective SP Hamiltonian (Eq.~(\ref{Hsp-quant})) and using the second-quantized representation for the corresponding dipole moments (Eqs.~(\ref{musp}) and (\ref{mux})), we arrive at the second term  of our Hamiltonian (Eq.~(\ref{Hsp-quant2})) with the exciton-SP coupling parmaetr
\begin{equation}\label{eq:lambda}
\lambda_{nk}=e^{-ink}[\lambda_0+2\lambda_1\cos k+i 2\lambda_2\sin k]=e^{-ink}\Lambda_k. 
\end{equation}
This  term depends on the interaction energy between $n$-th emitter the nanorod it dresses and the nearest neighbor nanorods given by  
\begin{eqnarray}
\label{Lm0-sprd}
\lambda_0 &=& \frac{3 p_{sp}\mu_{qe}}{8\pi\varepsilon_0 f_0^3}({\bf e}_{qe}\cdot\bm T(0,d)), 
\\
\label{Lm12-sprd}
\lambda_{1,2} &=& \frac{3p_{sp}\mu_{qe}}{16\pi\varepsilon_0 f_0^3}
\left[({\bf e}_{qe}\cdot\bm T(-a,d))
\right.  \pm \left.
({\bf e}_{qe}\cdot\bm T(a,d))\right],
\end{eqnarray}
respectively. Here, $a$ is distance between the centers of nearest neighbor nanords and $d$ is distance between the centers of a nanorod and the the quantum emitter sitting on top of it.  
Provided $f_0\ll r_{1,2}$, the coupling constants given by Eqs.~(\ref{Lm0-sprd}), and (\ref{Lm12-sprd}), acquire the following dipole-dipole interaction form 
\begin{eqnarray}
\label{Lm0}
	\lambda_0^{dd} &=& \frac{p_{sp}\mu_{qe}}{4\pi\varepsilon_0 d^3}\left[({\bf e}_{sp}\cdot {\bf e}_{qe})-3({\bf e}_{sp}\cdot {\bm y})
	({\bf e}_{qe}\cdot {\bm y})\right], 
\\\nonumber
\label{Lm1}
	\lambda_1^{dd} &=& \frac{p_{sp}\mu_{qe}}{4\pi\varepsilon_0 \left(a^2+d^2\right)^{3/2}}
\\\nonumber&\times&
	\left[({\bf e}_{sp}\cdot {\bf e}_{qe})-3\cos^2\alpha~({\bf e}_{sp}\cdot {\bm x})
	({\bf e}_{qe}\cdot {\bm x})
\right.\\\nonumber &-&\left.
	3\sin^2\alpha~({\bf e}_{sp}\cdot {\bm y})
	({\bf e}_{qe}\cdot {\bm y})\right],
\\\label{Lm2}
	\lambda_2^{dd} &=& \frac{3 p_{sp}\mu_{qe}}{8\pi\varepsilon_0 \left(a^2+d^2\right)^{3/2}}\sin 2\alpha
\\\nonumber&\times&
	\left[({\bf e}_{sp}\cdot {\bm x})
		({\bf e}_{qe}\cdot {\bm y})
	+({\bf e}_{sp}\cdot {\bm y})
	({\bf e}_{qe}\cdot {\bm x})\right],
\end{eqnarray}
where $\sin\alpha=d/(a^2+d^2)^{1/2}$. Alternatively, Eqs.~(\ref{Lm0})--(\ref{Lm2}) can be obtained from Eqs.~(\ref{Hsp-quant}) and (\ref{En-local-dipole}).

\begin{figure*}[t]
\subfigure[]{\includegraphics[width=0.32\textwidth]{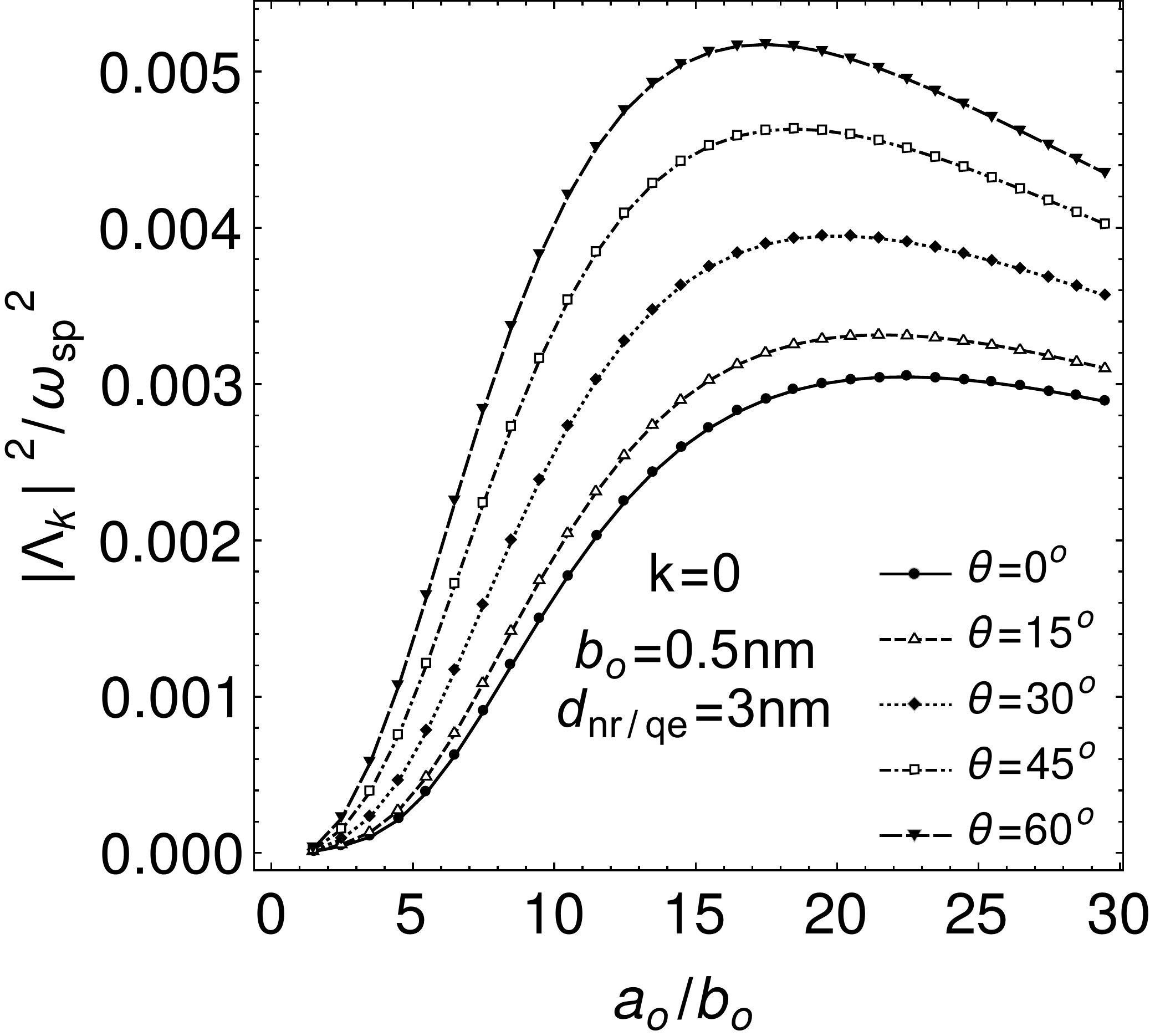}} 
\subfigure[]{\includegraphics[width=0.32\textwidth]{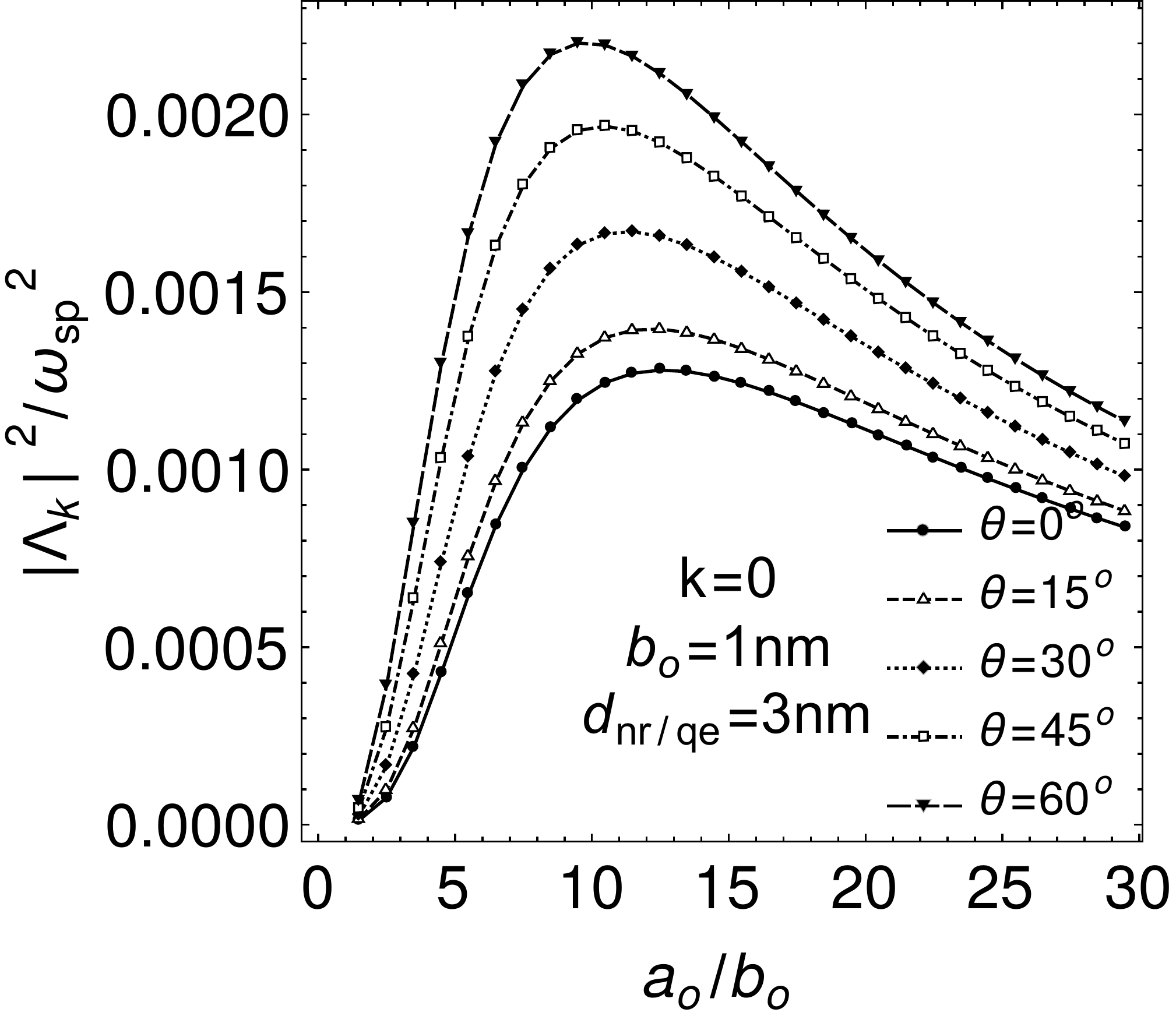}} 
\subfigure[]{\includegraphics[width=0.32\textwidth]{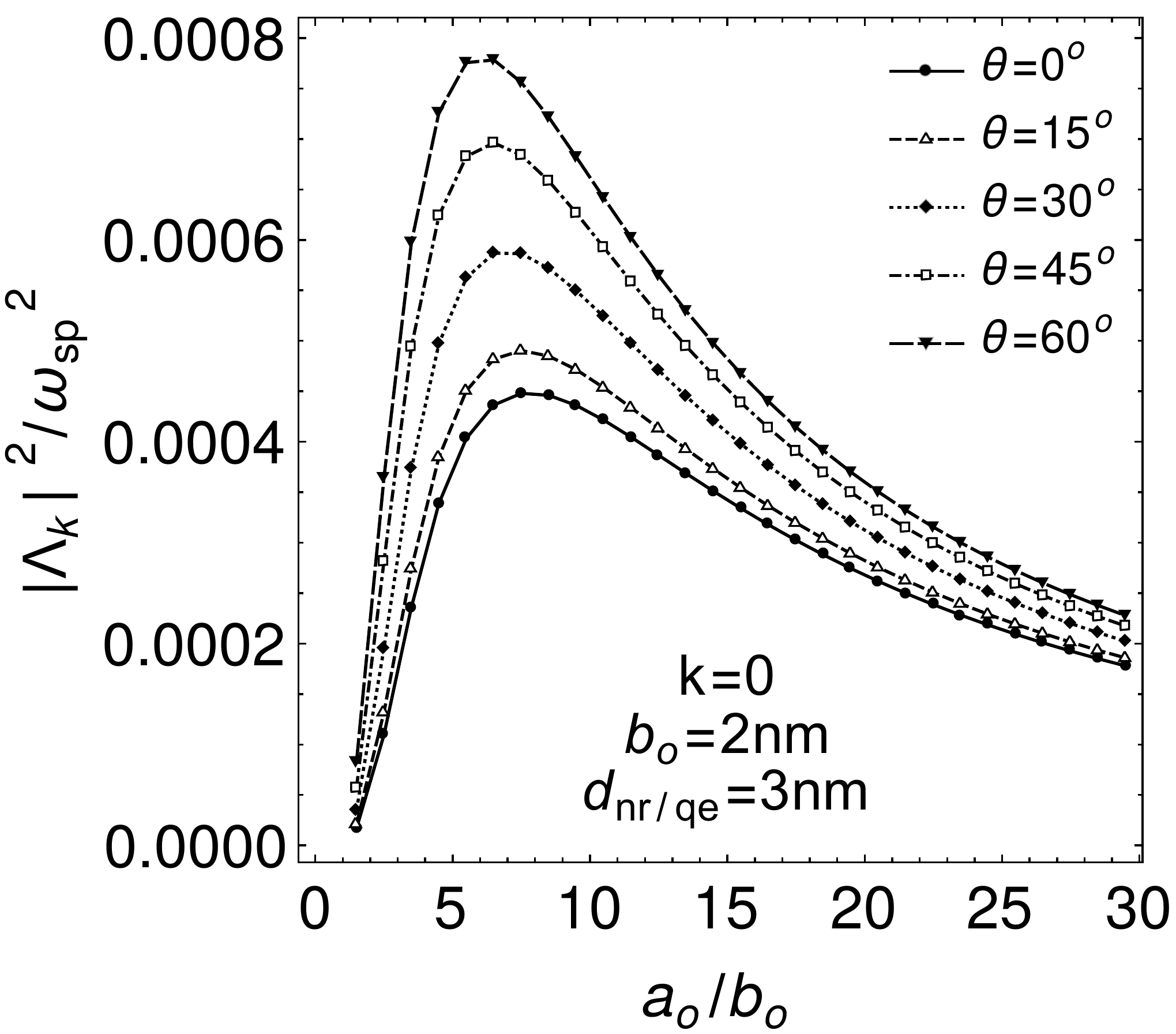}} 
\caption{Squared absolute value of the the exciton-SP coupling, $|\Lambda_k|^2$, calculated for various values of the nanorod aspect ratio $a_0/b_0$, width $b_0$, and tilt-angle $\theta$ at $k=0$ in units of the SP frequency, $\omega_{sp}$. Distance between each quantum emitter and the nanorod it dresses is set to $d=3$~nm.}
\label{Biglambdas}
\end{figure*}

In Fig.~\ref{Biglambdas} we show the absolute value of the the exciton-SP coupling, $|\Lambda_k|^2$, calculated according to Eqs.~(\ref{eq:lambda})--(\ref{Lm12-sprd}) for various values of the nanorod aspect ratio $a_0/b_0$, width $b_0$, and tilt-angle $\theta$ at $k=0$ in units of the SP frequency, $\omega_{sp}$.  The results indicate that the highest couplings are achieved with thin ($b_{o} =$ 0.5~nm)  nanorods  of aspect ratios between 10-15 and alignment defined by $\theta = 60^{\circ}$. The use of approximate Eqs.~(\ref{Lm0}) and (\ref{Lm2}) (not shown) does not bring significant deviations from the results in Fig.~\ref{Biglambdas}.

\subsection{Quantum emitter Hamiltonian}
\label{Sec:HQE}

We now focus on the last term in Eq.~(\ref{Hsp-quant2}) describing the quantum emitters and their coupling to each other resulting in the delocalized exciton states. Our exciton model for this is identical to the so-called XY model describing a linear chain of spins coupled to a transverse magnetic field with the Hamiltonian  
\begin{eqnarray}
\label{HX-def}
\hat H_{XY} =  \frac{\omega_{qe}}{2} \sum_{n}\hat\sigma_{n}^{z}+\frac{J_{qe}}{2}\sum_{n}\left(\hat\sigma_n^+\hat\sigma_{n+1}^-+\hat\sigma_{n+1}^{+}\hat\sigma_n^-\right).
\end{eqnarray}
Here, $\omega_{qe}$ is the two-level system transition energy. The nearest neighbor interaction between the quantum emitters can be well represented by the dipole-dipole interaction energy
\begin{eqnarray}
\label{Jsp-qe}
J_{qe}= \frac{\mu_{qe}^2}{4\pi\varepsilon_0 a^3}\left(1-3\left({\bf e}_{qe}\cdot{\bm x}\right)^2\right),
\end{eqnarray}
where $\mu_{qe}$ is the quantum emitter transition dipole and $a$ is the distance between neighboring emitters.

Since the model corresponds to the free-propagation of excitons on a lattice, one is tempted to immediately diagonalize $\hat H_{XY}$.
However, all raising/lowering operators belonging to different sites commute, $[\hat\sigma_{i},\hat\sigma_{j} ] = 0$ for $i\ne j$.  The Pauli principle requires that they anti-commute between sites.  Consequently, the $\hat\sigma_{i}^{\pm}=(\hat\sigma^x_{i}\pm\hat\sigma^-_{i})/\sqrt{2}$ operators as defined above describe {\em neither} fermions nor bosons and are inconvenient for describing a many-body system. This can be resolved  defining a new set of operators, (Holstein-Primakoff transformation)
\begin{eqnarray}
\hat\sigma^z_i &=& 2\hat b_i^\dag\hat b_i -1,
\\
\hat\sigma^+_i &=&\sqrt{2}\hat b_i^\dag\left(1-\hat b_i^\dag\hat b_i\right)^{1/2} 
\\
\hat\sigma^-_i &=&\sqrt{2}\left(1-\hat b_i^\dag\hat b_i\right)^{1/2}\hat b_i
\end{eqnarray}
where the $\hat b_i$'s do obey the canonical Bose algebra  $[\hat b_{i},\hat b_{j}^{\dagger}] = \delta_{ij}$. This transformation preserves spin operator commutation relation $[\hat\sigma_{i}^{+},\hat\sigma_{j}^{-} ]=2\delta_{ij}\hat\sigma^{z}_{i}$ and can be used to describe
the propagation of the quantum emitter excitons using the identity
\begin{eqnarray}
\frac{1}{2}\sum_{n=1}^{N} \left(\hat \sigma^{+}_{n}\hat\sigma^{-}_{n+1} + \hat\sigma^{+}_{n}\hat\sigma^{-}_{n+1}\right) &=& 
\sum_{n=1}^{N}
 \left(\hat b_{n}^{\dagger}\hat b_{n+1} + \hat b_{n+1}^{\dagger}\hat b_{n}\right) \nonumber \\
 &+& {\cal O}(1/N)
\end{eqnarray}
We introduce a small ${\cal O}(1/N)$ error associated with the terms operator terms higher than the second order. 
It is also important to note that the quantum emitter Hamiltonian explicitly preserves the total number of 
bosons (excitations) in the system viz. ${\hat N}_{b} = \sum_{n}\hat b_{n}^{\dagger}\hat b_{n}$ and $[\hat H_{XY},{\hat N}_{b}] = 0$. 
As a result, we can write total spin operator $S_{z} = \sum_{i}\hat\sigma_{i}^{z} = 2{\hat N}_{b}-N,
$ and observe that each spin-less excitation created by $\hat b^{\dagger}_{i}$ carries an $S_{z}\omega_{qe}/2 = \omega_{qe}$ quantum of energy.  

The quantum-emitter terms can now be brought into diagonal form by defining the transformation
\begin{eqnarray}
\hat H_{x} = \sum_{ij}\hat b_{i}^{\dagger}H_{ij}\hat b_{j}
\end{eqnarray}
where
\begin{eqnarray}
H_{ij} = \omega_{qe}\delta_{ij} + J_{qe}(\delta_{i,j+1} + \delta_{j,i+1})
\end{eqnarray}
Introducing the Fourier expansion of the Fermion operators
\begin{eqnarray}
\hat b_{n} &=& \frac{1}{\sqrt{N}}\sum_{k}e^{ik n}\hat b_{k}
\\
\hat b_{n}^{\dagger} &=&\frac{1}{\sqrt{N}}\sum_{k}e^{-ik n}\hat b^{\dagger}_{k},
\end{eqnarray}
where $k =-\pi,\dots,-\pi$, and $\hat b_{k}$'s are canonical boson operators describing delocalized exciton states with momentum $k$, we thus arrive at the diagonal {\em exciton} Hamiltonian 
\begin{eqnarray}
\hat H_{x} = \sum_{k}\omega_x(k)\hat b_{k}^{\dagger}\hat b_{k}
\end{eqnarray}
with the energy dispertion  
\begin{eqnarray}
\label{X-disp}
\Omega_x(k)=\omega_{qe}+2J_{qe}\cos(k). 
\end{eqnarray}
Similar to SP (Eq.~(\ref{wspk})), the position of the quantum emitter band edge within the Brillouin zone depends on the sign of $J_{qe}$. Since we keep orientation of the quantum emitter and SP transition dipoles parallel the sign of $J_{qe}$ has similar angular dependance as that shown for the SP in Fig.~\ref{Fig-Jvalues}c.   

\begin{widetext}
Pulling everything together, our model Hamiltonian (Eq.~(\ref{Hsp-quant2}) takes the form of a series of independent boson terms 
\begin{equation}\label{k_ham}
  \begin{split}
\hat H_{SP} &= 
\sum_{k}\left\{\Omega_{sp}(k)\hat\psi_{k}^{\dagger}\hat\psi_{k} +  \Omega_x(k) \hat b_{k}^{\dagger} \hat b_{k}
+
\left(\hat b_{k}^{\dagger}+\hat b_{k}\right)\left(\Lambda_{k}^{*}\hat\psi_{k}^{\dagger}+\Lambda_{k}\hat\psi_{k}\right)\right\}.
  \end{split}
\end{equation}
Note that this is similar to the Dicke model for the coupling between a photon cavity mode and an ensemble of spins which has 
been widely used in atomic and molecular physics.  However, in this case, each cavity (i.e., SP) mode, $\hat\psi_{k}$, is coupled to a single exciton mode $\hat s_{k}$ rather than to an ensemble of independent 2-state atoms.

\section{Quantum Symmetry Breaking}
\label{Sec:QSB}

Having defined our system by the Hamiltonian given in Eq.~(\ref{k_ham}), we diagonalize this Hamiltonian . For this purpose we adopt the rotating wave approximation (RWA) 
\begin{eqnarray}
\label{H-total-RWA}
\hat H_{RWA} &=&\sum_{k}\left\{
\Omega_{sp}(k)\hat\psi_{k}^{\dagger}\hat\psi_{k} + \Omega_x(k) \hat b_{k}^{\dagger} \hat b_{k}
+(\Lambda^*_{k}\hat\psi_{k}^{\dagger}\hat b_{k}+\Lambda_{k}\hat b_{k}^{\dagger}\hat\psi_{k})
\right\}.
\end{eqnarray}
To eliminate the exciton-SP coupling, the following polariton transformation for the SP and photon operators is adopted 
\begin{eqnarray}\label{alpha-trf}
\hat\psi_{k} &=& u_{k} \hat\Psi_{+}(k) + v_{k} \hat \Psi_{-}(k) \\
\hat b_{k} &=&   -v_{k} \hat \Psi_{+}(k)+u_{k} \hat\Psi_{-}(k) .
\end{eqnarray}
The unitary transformation matrix elements eliminating the desired term read
\begin{eqnarray}
\label{cosak}
u_{k} &=& \frac{\Omega_+(k)-\Omega_{sp}(k)}{\sqrt{\left(\Omega_+(k)-\Omega_{sp}(k)\right)^2+|\Lambda_k|^2}},
\\\label{sinak}
v_{k} &=& \frac{\Lambda_k}{\sqrt{\left(\Omega_+(k)-\Omega_{sp}(k)\right)^2+|\Lambda_k|^2}},
\end{eqnarray}
and result in the following transformed Hamiltonian 
\begin{eqnarray}
\label{H-plrt}
\hat H_{\pm} &=& \sum_{k} \left\{\Omega_{+}(k) \hat\Psi_{+}(k)^{\dagger}\hat\Psi_{+}(k) 
 + \Omega_{-}(k) \hat \Psi_{-}(k)^{\dagger}\hat\Psi_{-}(k)
\right\}
\end{eqnarray}
where the Bose operators $\hat\Psi_{+}(k)$ and $\hat\Psi_{-}(k)$ naturally describe the quasiparticles associated with the upper, $\Omega_+(k)$, and lower, $\Omega_-(k)$, polariton branches, respectively. The associated polariton energies are
\begin{eqnarray}
\label{OmgPM-k}
\Omega_{\pm}(k) &=&  \frac{\Omega_{ph}(k) + \Omega_{sp}(k)}{2} \pm 
\sqrt{\left(\frac{\Omega_{ph}(k)-\Omega_{sp}(k)}{2}\right)^{2} + |\Lambda_{k}|^{2}}.
\end{eqnarray}
  
\begin{figure*}
\begin{center}
\includegraphics[width=0.9\columnwidth]{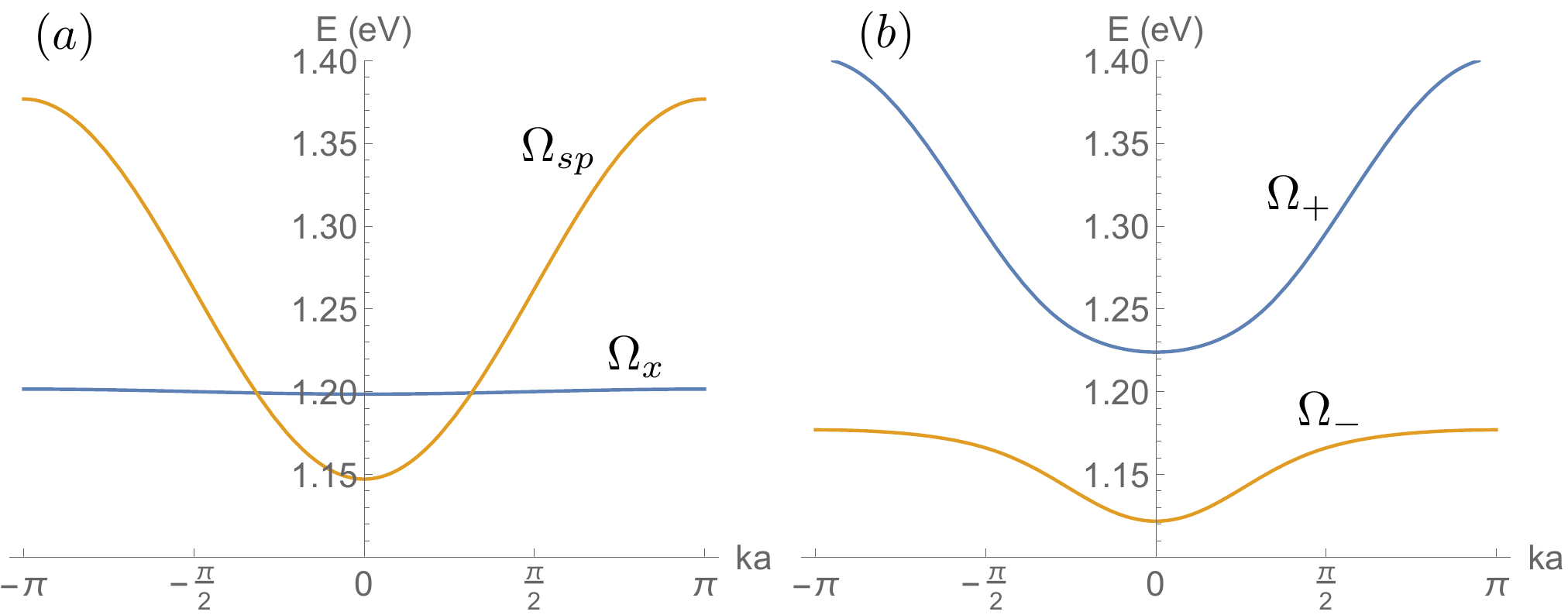}
\end{center}
\caption{Non-interacting SP, $\Omega_{sp}$, and exciton $\Omega_k$ dispersion curves calculated according to Eqs.~(\ref{wspk}) and (\ref{X-disp}), respectively. The following geometric parameters for the array are adopted in the calculations $a_o=10$~nm, $b_o=1$~nm, $\theta=0$, and $a=20$~nm. (b) Upper, $\Omega_+$, and lower, $\Omega_-$, polariton branches (Eq.~(\ref{OmgPM-k})) formed after exciton-SP interaction, $\Lambda_k$ (Eqs.~(\ref{eq:lambda})--(\ref{Lm12-sprd})) is taken into account for the parameters adopted in panel (a).  }
\label{Fig-polarts}
\end{figure*}

In Fig.~\ref{Fig-polarts}a, we plot non-interacting SP, $\Omega_{sp}(k)$, and exciton, $\Omega_k(k)$, dispersion curves. Turning on interaction $\Lambda_k$ splits the curves at the crossing points into polariton branches $\Omega_{\pm}$ as shown in panel b. These are analogous to the upper and lower polariton branches except that both the SP and excitonic modes are described by $\cos(k)$ dispersion terms. Comparing panels b with a, one can conclude that in the vicinity of $k=0$ the upper polariton branch, $\Omega_+(k)$, represents the exciton states dressed by SP and the lower one, $\Omega_-(k)$, the SP states dressed by excitons. The state dressing shows up as the energy-shift of the $\Omega_\pm(k=0)$ compared to the energies of  $\Omega_{sp}(k=0)$ and $\Omega_{x}(k=0)$.

\end{widetext}

At this point we focus on the SP-like, $\Omega_-(k)$ polariton branch and ask whether or not a 1D system described by a $\cos(k)$ dispersion can undergo a quantum phase transition to form a Bose-Einstein condensate (BEC). In general, the BEC transition is forbidden by symmetry for 1 and 2 dimensional systems. The reason for this can be traced to a non-removable singularity at $k = 0$ when integrating the number density over all momentum states and is the basis of the Mermin-Wagner theorem. However, for interacting systems and systems in potential wells, BEC is possible even in 1 and 2 dimensions. To determine whether or not such transitions are possible in this system, we consider the average population number for a weakly interacting Bose ensemble and compare the total ground state population to the total excited population at finite temperature. Let us introduce partition function
\begin{eqnarray}
\Xi = {\rm Tr}\left[e^{-\beta (\tilde H-\mu\hat N)}\right]
\end{eqnarray}
where $\mu$ is the chemical potential and $\hat N$ is the number of bosons in the system. 
We shall assume at this point that the system is open and that the total number of bosons in the system 
is allowed to fluctuate about some average value.  
Writing the free energy as $-\beta A =  \log \Xi$, we compute the occupation numbers for the lower 
polariton curve as 
\begin{eqnarray}
N  &=& \left( \frac{\partial A}{\partial \mu} \right)_{T,V}  \\
&=& \sum_{k = -\pi}^{\pi} \frac{\lambda e^{-\beta \Omega_{-}(k)}}{1 - \lambda e^{-\beta \Omega_{-}(k)}} \\
&=&  \frac{\lambda}{1-\lambda} +  \sum_{k \ne 0} \frac{\lambda e^{-\beta \Omega_{-}(k)}}{1 - \lambda e^{-\beta \Omega_{-}(k)}} \\
&=& N_{o} + N_{ex}
\end{eqnarray}
Here, we define the fugacity, $\lambda \in [0,1)$, such that the energy origin is at the bottom of the $\Omega_{-}(k)$ dispersion curve and $\beta = 1/k_{B}T$  and the sum is taken over the first Brillouin zone. The first term, thus represents the number of particles in the lowest energy state while the second sum reflects the number of particles in the excited states. BEC occurs when at finite temperature and $\lambda < 1$ the ground state population diverges logarithmically while the excited state population remains finite.


\begin{figure}[t]
\begin{center}
\includegraphics[width=0.80\columnwidth]{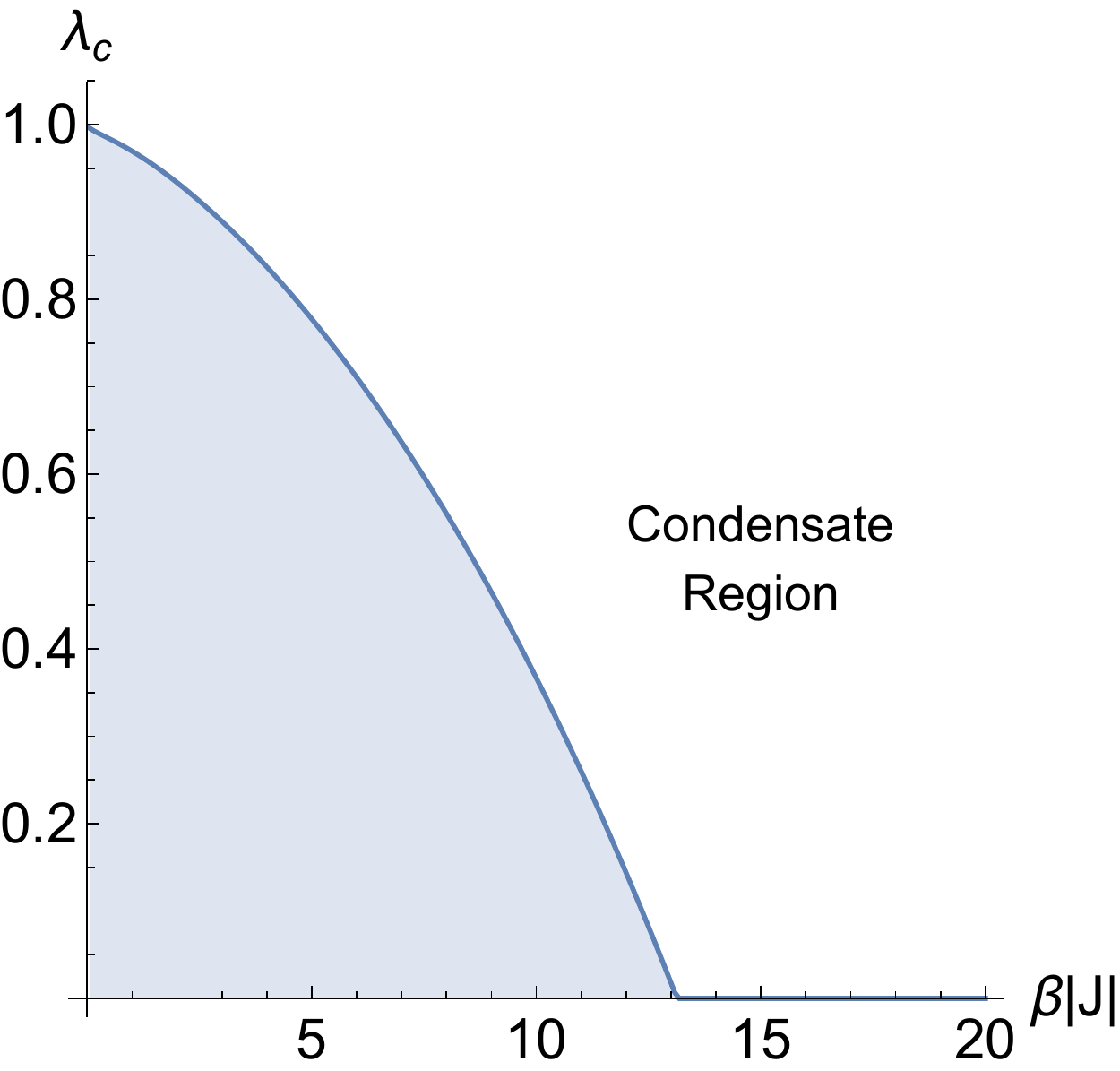}
\end{center}
\caption{Phase diagram for ideal Bose-gas in a 1D band. Here we take the scaled temperature $\beta |J|$
 as an order parameter and $\lambda_{c}$ as  the critical fugacity.
  }\label{BECPhaseDiagram}
\end{figure}

To a good approximation, close to $k=0$,  $\Omega_{-}(k) \approx |J|(1-\cos(k))/2 = |J|\sin^{2}(k/2)$. Thus, the integral over $k$ can be performed exactly producing an expression for the number density of particles in the excited states.
\begin{eqnarray}
\frac{N_{ex}}{V} &= &
2\int_{0^{+}}^{\pi}dk \frac{\lambda e^{-a \sin^{2}(k/2)}}{1-\lambda e^{-a \sin^{2}(k/2)}}  \nonumber \\
&=& \sum_{n=1}^{\infty}\lambda^{n} \left[2\int_{0^{+}}^{\pi}dk e^{-a n\sin^{2}(k/2)} \right] \nonumber \\
&=&
2 \pi \sum_{n=1}^{\infty}\lambda^{n}e^{-a n/2}I_{0}(an/2)
\end{eqnarray}
where $a = \beta|J|$ and $I_{0}(x)$ is the modified Bessel function of the first kind. The final sum can be evaluated numerically and rapidly converges for a given finite value of $\beta |J| > 0$ and $0 \le \lambda < 1$. At some critical value of $\lambda$ and finite temperature, $N_{ex} < N_{0}$ implying that under these conditions, the population will collapse to the lowest wave-vector state with decreasing temperature. In Fig. 5 we present a phase diagram showing the critical fugacity vs. $\beta |J|$. A low chemical potential corresponds to larger values of fugacity. Thus, for a given value of the chemical potential for the system, we can predict the critical temperature for the on-set of the BEC transition. This analysis implies that that in the dressed nanorod system we describe, one should be able to see the on-sent of the BEC transition at finite temperature and experimentally accessible excitation densities. In Fig.~\ref{BECPhaseDiagram} we show the phase diagram for the a Bose gas in a 1D band taking $\beta |J|$ as a scaled (inverse) temperature and $\lambda_{c}$ as the critical fugacity.  At low temperatures, the system is expected to become super-radiant for all temperatures $T^{-1} \gtrsim 13\beta |J|$. 

\section{Conclusion} 
\label{Sec:Conc}
We propose a model for interacting collective SP modes and exciton states in an array of metal nanorods dressed by quantum emitters. Starting with a second-quantized picture of the anisotropic SP response, we construct a model Hamiltonian and explicitly compute the coupling required parameters. The final model can be solved exactly, resulting in upper and lower polariton  dispersion curves. Moreover, we show that the model admits a BEC transition to create a condensate of SP-excitonic states. Our work suggests a set of  material parameters and nanorod alignments that can be optimized to achieve this condition under currently achievable laboratory and synthetic conditions.\cite{HollingsworthJ:2015}

\section{Acknowledgments.} The work at the University of Houston was funded in part by the National Science Foundation (CHE-1362006) and the Robert A. Welch Foundation (E-1337). The work at Los Alamos National Lab  was funded by the Los Alamos Directed Research and Development (LDRD) funds.

%

\end{document}